\documentclass[prb,aps,twocolumn,amsmath,amssymb,floatfix,showpacs,
superscriptaddress]{revtex4}

\usepackage[dvips]{graphics}
\usepackage{color}
\definecolor{dred}{rgb}{0,0,0.6}

\begin{document}

\title{\textcolor{dred}{Circulating current in 1D Hubbard rings with 
long-range hopping: Comparison between exact diagonalization method and 
mean-field approach}}

\author{Madhumita Saha}

\email{madhumitasaha91@gmail.com}

\affiliation{Physics and Applied Mathematics Unit, Indian Statistical
Institute, 203 Barrackpore Trunk Road, Kolkata-700 108, India}

\author{Santanu K. Maiti}

\email{santanu.maiti@isical.ac.in}

\affiliation{Physics and Applied Mathematics Unit, Indian Statistical
Institute, 203 Barrackpore Trunk Road, Kolkata-700 108, India}

\begin{abstract}

The interplay between Hubbard interaction, long-range hopping and disorder
on persistent current in a mesoscopic one-dimensional conducting ring 
threaded by a magnetic flux $\phi$ is analyzed in detail. Two different 
methods, exact numerical diagonalization and Hartree-Fock mean field 
theory, are used to obtain numerical results from the many-body Hamiltonian. 
The current in a disordered ring gets enhanced as a result of electronic 
correlation and it becomes more significant when contributions from higher 
order hoppings, even if they are too small compared to nearest-neighbor 
hopping, are taken into account. Certainly this can be an interesting 
observation in the era of long-standing controversy between theoretical 
and experimental results of persistent current amplitudes. Along with 
these we also find half-flux quantum periodic current for some typical 
electron fillings and kink-like structures at different magnetic fluxes 
apart from $\phi=0$ and $\pm \phi_0/2$. The scaling behavior of current 
is also discussed for the sake of completeness of our present analysis.

\end{abstract}


\maketitle

\section{Introduction}

The phenomenon of persistent current was first established during $1980$'s
when B\"{u}ttiker and his group have shown~\cite{butt1} that a small 
conducting ring carries a net circular current in presence of an 
Aharonov-Bohm (AB) flux $\phi$, and it does not vanish even in presence 
of disorder. This is a pure quantum mechanical phase coherence phenomenon 
and a direct consequence of AB effect~\cite{imry1}.

The experimental verification of it came after few years when Levy and
co-workers did the pioneering experiment~\cite{levy} considering an ensemble 
of $10^7$ independent Cu rings. Later several other experiments~\cite{jari,
bir,chand,mailly1,mailly,blu} were carried out to confirm the existence of 
persistent current in different conducting loops. At the same time a
substantial amount of theoretical work~\cite{gefen,yang,cheu2,mont,bouz,
giam,san1,san2,madhu,ambe,schm1,schm2,bary} 
has also revealed several unique features along this direction. From the 
single ring 
measurement~\cite{levy} a period of $\phi_0$ ($=ch/e$, the elementary 
flux-quantum) current has been observed, while some other 
experiments~\cite{jari,ram} have shown both $\phi_0/2$ and $\phi_0$ periodic 
currents. This half flux-quantum ($\phi_0/2$) periodic persistent current 
is not well explained from theoretical calculations and certainly it demands
proper analysis. The another confliction between theoretical and 
experimental results is associated with the prediction of low-field
currents, viz, slope of the current with flux $\phi$ in the limit 
$\phi \rightarrow 0$. Some experiments~\cite{levy,jari} have shown 
diamagnetic (negative slope) nature, while paramagnetic (positive slope) 
response has been noticed from other experiments~\cite{chand}. These 
observations do not really match well with theoretical observations. 
Many attempts have been made along this line and it has been observed 
that~\cite{cheu2} the slope strongly depends on electronic filling, 
randomness, impurities, etc.

Finally, the most important controversy has been raised during the 
prediction of current amplitude. Only for nearly ballistic rings comparable
currents are obtained, while for diffusive rings the calculated current 
is much smaller and in some cases it is even less than two orders of
magnitude compared to the experimental observations. In order to resolve
this long-standing anomaly in diffusive rings, several attempts have been
made considering the effects of different parameters. It has been verified
that on-site Coulomb correlation plays a dominant role~\cite{mont,bouz,giam} 
to enhance current amplitude in a disordered ring, though the enhancement
is not sufficient to make it comparable to experimental observations. 
Later few theoretical models have been given considering the effect of 
higher order hopping integrals~\cite{san3,san4} in addition to the usual 
nearest-neighbor hopping (NNH) in favor of current enhancement upto a 
certain level. So naturally the question comes how the current becomes 
affected if one includes the effect of both on-site Coulomb interaction 
and long-range hopping. This part has not been discussed so far, to the 
best of our knowledge. 

In the present work we actually focus towards this direction. Here we
consider interacting rings with higher-order hopping integrals and 
determine energy eigenvalues by {\em exact numerical diagonalization} 
of many-body tight-binding Hamiltonian. In this method we restrict ourselves
to rings with few number of electrons, because of our computational
limitations, as dimension of the matrix raises rapidly with ring size $N$ 
and total number of electrons $N_e$. Different combinations of up and down 
spin electrons are taken into account and from our results we observe some 
similarities in current-flux characteristics for odd $N_e$ as well as even 
$N_e$, which certainly help us to generalize the results for larger ring 
size with higher electrons. However, it is always
good to have some definite results of large interacting rings to understand
the interplay between long-range hopping, disorder and e-e interaction. We 
thus require an appropriate technique to work out such a many-body problem. 
Here we use Hartree-Fock (HF) mean-field theory~\cite{kato,kam,sil,san5} 
to find energy eigenvalues and hence
persistent currents, and, compare these currents with exact exact numerical 
diagonalization analysis. Scrutinizing the results we show that persistent
current behaves almost identical in these two methods, but one can miss some 
important signatures of current at some typical fluxes, particularly in the 
limit $\phi \rightarrow 0$. This is an artifact of the mean-field 
calculations.

Several interesting features are obtained from our calculations. The 
current gets enhanced in a disordered ring as a result of Coulomb repulsion
and it becomes more effective when higher order hopping integrals are
included. Under this situation current becomes quite comparable to that
of a perfect ring. In addition we also discuss the appearance of half
flux-quantum periodicity for some particular electron fillings and some
kink-like structures at different fluxes apart from $\phi=0$ and 
$\pm \phi_0/2$. Finally, scaling behavior of current is also discussed,
for the sake of completeness.

The rest of the work is arranged as follows. In Sec. II we describe the 
ring model and theoretical procedure to construct the Hamiltonian matrix
and calculation of persistent current as a function of flux $\phi$. Later,
in Sec. III we discuss numerical results, and finally, a brief summary 
is given in Sec. IV.

\section{Model and Theoretical Formulation}

\subsection{Model and Tight-Binding Hamiltonian}

\begin{figure}[ht]
{\centering \resizebox*{4.75cm}{2.8cm}{\includegraphics{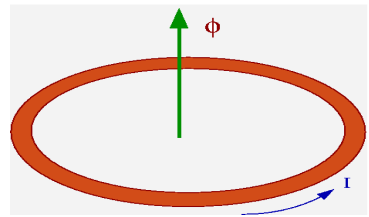}}\par}
\caption{(Color online). Schematic view of a mesoscopic ring threaded by 
an Aharonov-Bohm flux $\phi$. A net circulating current $I$ is established 
in the ring.}
\label{f1}
\end{figure}

The schematic diagram of our model is illustrated in Fig.~\ref{f1}, where
a mesoscopic ring is subjected to on-site Hubbard interaction and
long-range hoppings, viz, second-neighbor hopping (SNH) and 
third-neighbor hopping (TNH) including nearest-neighbor hopping. Under
the application of an AB flux $\phi$, a net circulating current is
established in the ring.

To describe this model we employ a tight-binding (TB) framework. For a
$N$-site ring the TB Hamiltonian looks like,
\begin{eqnarray}
\textbf{H} & = & \sum_{j,\sigma} \epsilon_j c_{j,\sigma}^{\dagger} 
c_{j,\sigma} + t \sum_{j,\sigma} \left[e^{i \theta_1} c_{j,\sigma}^{\dagger} 
c_{j+1,\sigma} \right. \nonumber \\
& & \left. + \, e^{-i \theta_1} c_{j+1,\sigma}^{\dagger} c_{j,\sigma}\right]
\nonumber\\
 & & + \, t_1 \sum_{j,\sigma} \left[e^{i \theta_2} c_{j,\sigma}^{\dagger} 
c_{j+2,\sigma} + e^{-i \theta_2} c_{j+2,\sigma}^{\dagger} c_{j,\sigma}\right]
\nonumber\\
 & & + \, t_2 \sum_{j,\sigma} \left[e^{i \theta_3} c_{j,\sigma}^{\dagger} 
c_{j+3,\sigma} + e^{-i \theta_3} c_{j+3,\sigma}^{\dagger} c_{j,\sigma}\right]
\nonumber\\
 & & + \, U \sum_j c_{j,\uparrow}^{\dagger} c_{j,\uparrow} 
c_{j,\downarrow}^{\dagger} c_{j,\downarrow} 
\label{eq1}
\end{eqnarray}
where $\epsilon_j$ gives the site energy for $j$th site and 
$c_{j,\sigma}^{\dagger}$ ($c_{j,\sigma}$) represents the creation 
(annihilation) operator. $t$ describes the nearest-neighbor hopping 
strength and due to this hopping a phase factor $\theta_1$ 
($=2\pi \phi/\phi_0$) is introduced into the Hamiltonian. Similarly, $t_1$
and $t_2$ denote the SNH and TNH integrals, respectively, with the 
associated phase factors $\theta_2=2\theta_1$ and $\theta_3=3\theta_1$.
The parameter $U$ measures the Hubbard interaction strength. Setting
$\epsilon_j=$ constant $\forall j$ we get a perfect ring and without loss
of generality we can fix it to zero, while for a disordered ring site 
energies are chosen {\em randomly} from a Box distribution function
of width $W$. 

\subsection{Method I: Exact Numerical Diagonalization}

Using the TB Hamiltonian, Eq.~\ref{eq1}, we construct matrix elements 
through the prescription $\textbf{H}_{mn} = 
\langle \psi_{m}|\textbf{H}|\psi_{n}\rangle$, where $|\psi_m\rangle$
and $|\psi_n\rangle$ are basis states associated with total number
of up ($N_{up}$) and down ($N_{dn}$) spin electrons in a ring. For example,
for a ring with two opposite spin electrons (viz, $N_{up}=N_{dn}=1$) the
state vectors are like:
$|\psi_{m}\rangle=c_{p\uparrow}^\dagger c_{q\downarrow}^\dagger |0\rangle$ 
and
$|\psi_{n}\rangle=c_{k\uparrow}^\dagger c_{l\downarrow}^\dagger |0\rangle$,
where $|0\rangle$ is the vacuum state. In a similar way, we can choose 
the basis states for any ring with different $N_{up}$ and $N_{dn}$ and
construct the required many-body Hamiltonian matrix. Diagonalizing this
matrix, we find energy eigenvalues and eventually determine persistent
current at absolute zero temperature from the relation
\begin{eqnarray}
I(\phi) & = & -\frac{\partial E_0(\phi)} {\partial \phi},
\label{equ3}
\end{eqnarray}
where $E_0(\phi)$ is the ground state energy. 

\subsection{Method II: Hartree-Fock Mean-Field Approach}

In this method the interacting Hamiltonian $\textbf{H}$ is decoupled into 
two non-interacting parts~\cite{sil,san5}, say, $\textbf{H}_{\uparrow}$ 
and $\textbf{H}_{\downarrow}$, associated with up and down spin electrons.
Diagonalizing these non-interacting Hamiltonians ($\textbf{H}_{\uparrow}$ 
and $\textbf{H}_{\downarrow}$), we find energy eigenvalues, and eventually,
calculate the ground state energy at absolute zero temperature for a
system containing $N_{up}$ and $N_{dn}$ spin electrons from the relation
\begin{equation}
E_0=\sum_{n=1}^{N_{up}} E_{n\uparrow} + \sum_{n=1}^{N_{dn}} E_{n\downarrow}
- \sum_{i=1}^N U \langle n_{i\uparrow} \rangle \langle n_{i\downarrow} \rangle
\label{eqnew} 
\end{equation}
where $E_{n\sigma}$'s are the energy eigenvalues and 
$\langle n_{i\sigma}\rangle=\langle c_{i\sigma}^{\dagger} c_{i\sigma}\rangle$.

Finally, persistent current is determined from the relation given in
Eq.~\ref{equ3}.

\section{Results and Discussion}

We present our results in two different sub-sections. In the first part
(sub-section A) we determine persistent current by evaluating ground state 
energy from {\em exact numerical diagonalization} of many-body Hamiltonian. 
Starting from a ring with two opposite spin electrons we consider upto 
six-electron system. For a fixed $N_e$ ($=N_{up} + N_{dn}$), different 
combinations of $N_{up}$ and $N_{dn}$ are taken into account to explore 
all the basic features of persistent current in interacting rings. While,
in the other part (sub-section B), persistent current is found out by
determining ground state energy from HF mean-field approach.

Throughout the analysis we set $t=-1\,$eV ($t_1$ and $t_2$ are variable 
and they are mentioned in appropriate places) and all the energies are 
scaled with respect to it. The current is measured in unit of $et/h$.

\subsection{Exact numerical diagonalization analysis}

\subsubsection{Impurity free rings}

Let us first concentrate on interacting rings without any impurity i.e.,
$\epsilon_j=0$ $\forall \; j$.

\vskip 0.25 cm
\noindent
\textbf{Case 1 - Ring with two electrons:} To have an interacting ring with
two electrons we consider one up and one down spin electron i.e.,
$N_{up}=N_{dn}=1$. For this configuration we select basis states as
$|\psi_m\rangle=c_{p\uparrow}^{\dagger} c_{q\downarrow}^{\dagger}|0\rangle$
and 
$|\psi_n\rangle=c_{k\uparrow}^{\dagger} c_{l\downarrow}^{\dagger}|0\rangle$
and the dimension of matrix becomes $N^2\times N^2$ for a $N$-site ring.

\begin{figure}[ht]
{\centering \resizebox*{8cm}{8cm}{\includegraphics{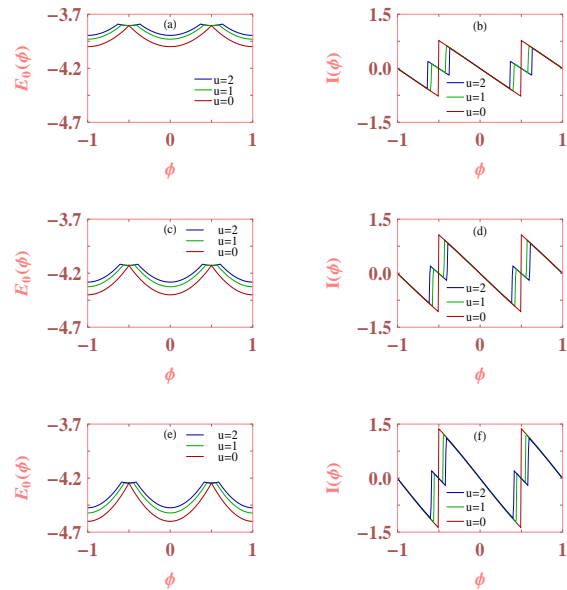}}\par}
\caption{(Color online). Dependence of ground state energy with flux $\phi$
and associated current for a $10$-site ring with two opposite spin electrons
for different values of $U$, where the $1$st, $2$nd and $3$rd rows correspond
to the ring with only NNH; NNH and SNH; and NNH, SNH and TNH integrals, 
respectively. Here we set $t_1=-0.1$ and $t_2=-0.05$.}
\label{f2}
\end{figure}

Figure~\ref{f2} displays the variation of ground state energy and 
corresponding persistent current with $\phi$ for a $10$-site ring having 
two opposite spin electrons for different values of correlation strength $U$.
In the $1$st row the results are shown when the ring is described with only
NNH integral, while in the $2$nd row we add the effect of SNH integral
and finally in the last row both SNH and TNH integrals are taken into 
account. The results are noteworthy. (i) The ground state energy exhibits
$U$-independent energy region across $\phi=\pm \phi_0/2$ and the width of 
this region gets increased with increasing $U$, while it decreases with 
the inclusion of higher order hopping integrals, as clearly seen by 
comparing the $E_0$-$\phi$ spectra given in Figs.~\ref{f2}(a), (c) and (e). 
This behavior is nicely reflected in current-flux characteristics yielding 
a kink-like structure. A sudden change in direction of current takes place
as a result of this kink, and interestingly we see that inside a kink
the slope of the current remains unchanged irrespective of $U$. (ii) For
a fixed correlation strength $U$, the slope of $E_0$-$\phi$ curve gets 
increased with higher order hopping which results a larger current. Here 
it is important to note that though TNH strength is much weaker than
SNH, it contributes significantly to raise the current which is visible
from the spectra given in Figs.~\ref{f2}(d) and (f). Certainly, more current
is expected in presence of additional higher order hopping, beyond TNH, 
though their strengths are vanishingly small. Physically this is easily
understandable since the incorporation of additional hopping allows 
electrons to hop further and thereby increases electronic current. 
(iii) A reduction of current is observed with increasing $U$ when all the 
other physical parameters describing the system remain unchanged. This is 
solely due to the repulsive nature of electron-electron (e-e) interaction. 
It suggests that lesser and lesser current is expected with increasing $U$,
though it eventually reaches to a finite value (not shown here) and never 
drops to zero since $N_e < N$, which we verify through our detailed 
numerical analysis.

\begin{figure}[ht]
{\centering \resizebox*{7cm}{5cm}{\includegraphics{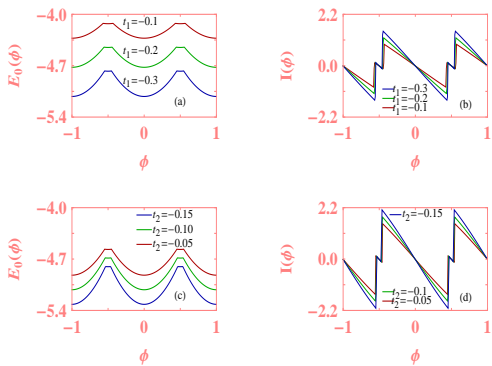}}\par}
\caption{(Color online). Variation of ground state energy and corresponding
current as function of $\phi$ for a $10$-site ring with two opposite spin
electrons considering $U=1$ for different values of $t_1$ and $t_2$. For
the $1$st row the ring is described with NNH and SNH, while in the $2$nd 
row the effect of TNH is also included where $t_1$ is fixed at $-0.2$.}
\label{f3}
\end{figure}

To have a deeper insight about the role of SNH and TNH integrals in 
Fig.~\ref{f3} we present the variation of ground state energy along with
persistent current as a function of $\phi$ considering $U=1$ for the
same ring size as taken in Fig.~\ref{f2} for some typical values of $t_1$
and $t_2$. From the spectra given in $1$st row we see that with increasing 
$t_1$ current gets enhanced 
following energy-flux diagram but it ($t_1$) reduces the kink size. 
Further enhancement of current can be also made possible with the 
inclusion of TNH (see Fig.~\ref{f3}(d)), and like $t_1$, $t_2$ is
also responsible for the reduction of kink size, though these kinks
do not vanish even for larger $t_1$ and $t_2$ as long as e-e interaction
is there. For this two-spin system we get only $\phi_0$ periodic current.

\vskip 0.25 cm
\noindent
\textbf{Case 2 - Ring with three electrons:} Now we consider a ring with one
down spin and two up spin electrons to have a three-electron ring system, and
here we choose $|\psi_1\rangle = c_{p\uparrow}^{\dagger} 
c_{q\uparrow}^{\dagger} c_{r\downarrow}^{\dagger} |0\rangle$ and 
$|\psi_2\rangle = c_{k\uparrow}^{\dagger} 
c_{l\uparrow}^{\dagger} c_{m\downarrow}^{\dagger} |0\rangle$. With this 
configuration the dimension of the matrix for a $N$-site ring becomes:
$N^2(N-1)/2 \times N^2(N-1)/2$. Two different cases are analyzed depending
on the filling factor those are as follows.

\begin{figure}[ht]
{\centering \resizebox*{6cm}{5cm}{\includegraphics{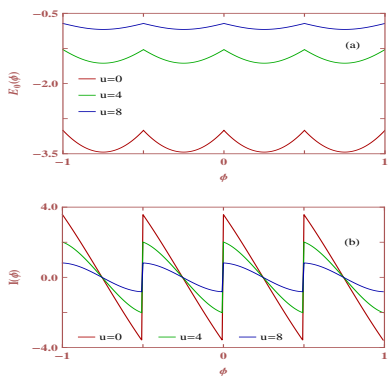}}\par}
\caption{(Color online). $E_0$-$\phi$ and corresponding $I$-$\phi$ 
characteristics of a $3$-electron ($N_{up}=2$ and $N_{dn}=1$) system in
the half-filled limit (viz, $N=3$). Here SNH and TNH are not applicable.}
\label{f4}
\end{figure}

\vskip 0.25 cm
\noindent
$\bullet$ \underline{Half-filled ring:} In Fig.~\ref{f4} we present the
variation of ground state energy $E_0$ and its corresponding persistent
current as a function of AB flux $\phi$ in the half-filled limit for 
three typical values of $U$. Two important features are observed.
(i) Ground state energy exhibits half-flux-quantum ($\phi_0/2$) periodicity
with $\phi$, and it is nicely reflected in the current-flux characteristics.
To illustrate this fact we can set the interaction strength $U$ to zero,
since it does not affect $\phi_0/2$ periodicity (which is shown from the
$E_0$-$\phi$ and $I$-$\phi$ spectra). Due to this simplification we can 
\begin{figure}[ht]
{\centering \resizebox*{8cm}{8cm}{\includegraphics{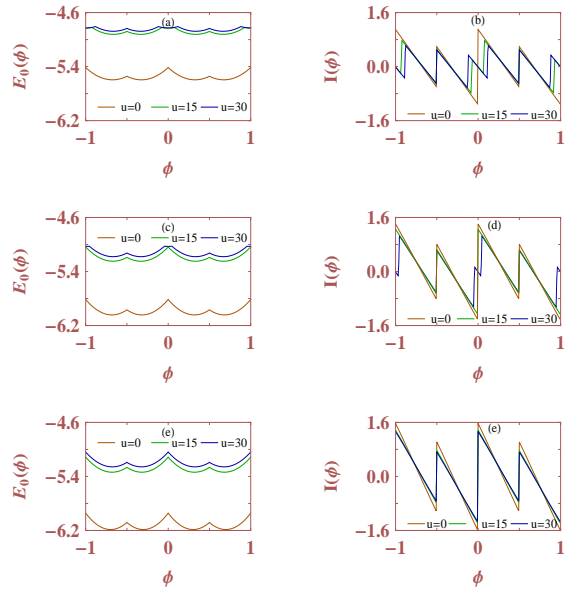}}\par}
\caption{(Color online). $E_0$-$\phi$ and $I$-$\phi$ curves for a $8$-site
ring with three electrons ($N_{up}=2$, $N_{dn}=1$), where the first, second
and third rows correspond to the ring with only NNH; NNH and SNH; NNH, SNH 
and TNH integrals, respectively. The other parameters are: $t_1=-0.1$ and
$t_2=-0.05$.}
\label{f5}
\end{figure}
handle the problem very easily as it becomes now a non-interacting one.
For $N=3$, three energy eigenstates are obtained where each of these states
can have maximum two opposite spin electrons. Therefore, ground state energy
is obtained by placing two opposite spin electrons in the lowest 
non-interacting energy level and the third electron (up spin) in the first
excited level. Quite interestingly, we observe that all such distinct energy
levels have $\phi_0$ flux-quantum periodicity. But as long as the highest
accessible energy level (here it is $1$st excited state) contains a 
{\em single electron}, instead of two, the net ground state energy of the
system gets strange $\phi_0/2$ periodicity. This peculiar $\phi_0/2$
periodicity also remains unchanged even in presence of $U$, which we claim 
from our results. (ii) No kink-like structure appears in $I$-$\phi$ curves
which implies that at half-filling no $U$-independent state contributes
to the lowest energy for any flux window.

For such filling, current decreases rapidly with the correlation strength
$U$, as there is no empty site.

\vskip 0.25 cm
\noindent
$\bullet$ \underline{Non-half-filled ring:} The results of a three-electron
non-half-filled ring ($N=8$) are shown in Fig.~\ref{f5} which exhibit some
interesting patterns unlike the half-filled case. (i) From the spectra it
is shown that the $\phi_0/2$ periodicity disappears and the current, 
associated with ground state energy, regains its usual $\phi_0$ periodicity.
(ii) The kink-like structure, associated with $U$-independent states, appears
in persistent current across $\phi=0$ at sufficiently large value of $U$, and
the width of such kink decreases with the inclusion of higher order hopping
integrals like SNH and TNH. In this context it is important to note that for
two-electron case, kink appears as long as the interaction is turned on.
The other features viz, the reduction of current with $U$ and its enhancement
with higher order hopping integrals remain same as discussed in a 
two-electron ring system.

\vskip 0.25 cm
\noindent
\textbf{Case 3 - Ring with four electrons:} For four-electron system we 
consider two distinct configurations depending on $N_{up}$ and $N_{dn}$. 
In one configuration we set $N_{up}=3$ and $N_{dn}=1$, while for the other 
we choose $N_{up}=N_{dn}=2$.

\vskip 0.25 cm
\noindent
$\blacksquare$ Configuration I -- Ring with $N_{up}=3$ and $N_{dn}=1$: \\
The dimension of the Hamiltonian matrix for this setup becomes 
$(N^2/6)(N-1)(N-2) \times (N^2/6)(N-1)(N-2)$, where the basis states are:
$|\psi_1\rangle = c_{p\uparrow}^{\dagger} c_{q\uparrow}^{\dagger} 
c_{r\uparrow}^{\dagger} c_{s\downarrow}^{\dagger} |0\rangle$ and 
$|\psi_2\rangle = c_{k\uparrow}^{\dagger} c_{l\uparrow}^{\dagger} 
c_{m\uparrow}^{\dagger} c_{n\downarrow}^{\dagger} |0\rangle$. 

\begin{figure}[ht]
{\centering \resizebox*{6cm}{5cm}{\includegraphics{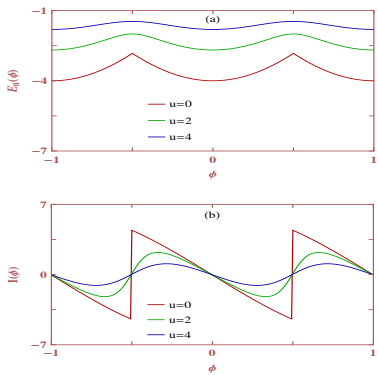}}\par}
\caption{(Color online). $E_0$-$\phi$ and $I$-$\phi$ curves of a four-site
ring ($N=4$) with $N_e=4$ ($N_{up}=3$ and $N_{dn}=1$) for different values 
of $U$ considering NNH only.}
\label{f6}
\end{figure}

The characteristics features of ground state energy and corresponding 
current for a half-filled ring with four electrons ($N_{up}=3$ and 
$N_{dn}=1$) are shown in Fig.~\ref{f6}. At $U=0$, a sharp transition
occurs in persistent current around $\phi=\pm \phi_0/2$, while for 
finite $U$ the current gets reduced and varies continuously with 
$\phi$. This continuous-like variation is not quite trivial as obtained 
in traditional disordered case. While, the reduction of current with $U$ 
is anticipated easily for this half-filled case.

\begin{figure}[ht]
{\centering \resizebox*{8cm}{8cm}{\includegraphics{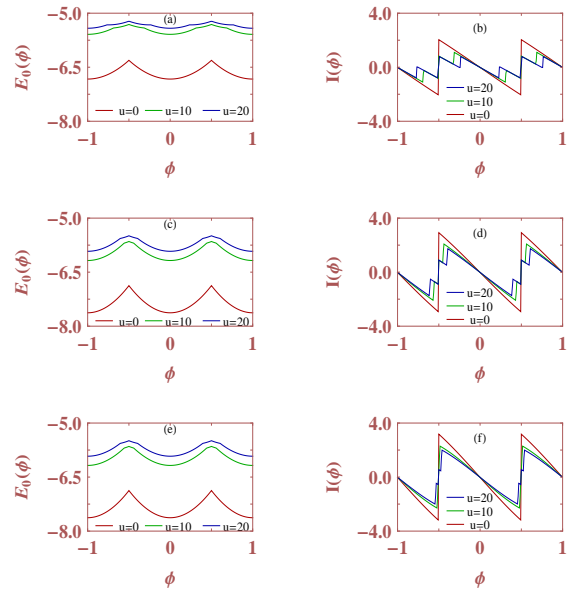}}\par}
\caption{(Color online). Energy and current spectra for a $8$-site ring
with $N_{up}=3$ and $N_{dn}=1$ for some typical values of $U$, where the
first, second and third rows correspond to the ring with only NNH; NNH and
SNH; NNH, SNH and TNH integrals, respectively. Here we choose $t_1=-0.1$
and $t_2=-0.05$.}
\label{f7}
\end{figure}

Away from half-filling some kink-like structures are visible in 
$I$-$\phi$ spectra reflecting the $E_0$-$\phi$ curves, and the most
crucial point is that these kinks are generated as a result of 
$U$-dependent states, not like $U$-independent states as described
earlier. The results are shown in Fig.~\ref{f7} for a $8$-site ring
considering the effects of higher order hopping (SNH and TNH) integrals 
for different values of $U$. The kink appears for a sufficiently large
value of $U$ and its width gradually decreases as a result of higher
order hopping.

\vskip 0.25 cm
\noindent
$\blacksquare$ Configuration II -- Ring with $N_{up}=2$ and $N_{dn}=2$: \\
For this configuration we choose many-body basis states as:
$|\psi_{1}\rangle=c_{p\uparrow}^\dagger c_{q\uparrow}^\dagger c_{r\downarrow}^
\dagger c_{s\downarrow}^\dagger |0\rangle$ and
$|\psi_{2}\rangle=c_{k\uparrow}^\dagger c_{l\uparrow}^\dagger c_{m\downarrow}^
\dagger c_{n\downarrow}^\dagger |0\rangle$. Here the dimension of the 
Hamiltonian matrix becomes $(N^2/4) (N-1)^2 \times (N^2/4) (N-1)^2$, and,
it is higher compared to the Configuration I, i.e., ring with three up
and one down spin electrons.

\begin{figure}[ht]
{\centering \resizebox*{6cm}{5cm}{\includegraphics{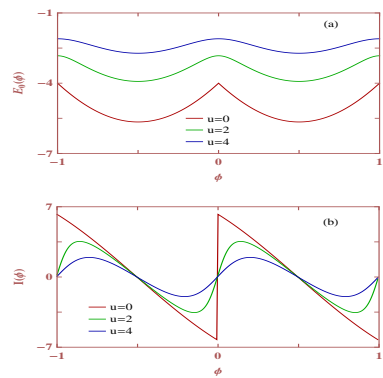}}\par}
\caption{(Color online). Variation of ground state energy and corresponding
current for a four-site ring with $N_{up}=2$ and $N_{dn}=2$ for different
strengths of $U$ considering NNH only.}
\label{f8}
\end{figure}

The results for a half-filled ring are displayed in Fig.~\ref{f8}. It is
shown that in absence of electronic correlation ($U=0$), current exhibits
a sharp transition similar to that as obtained in the previous $4$-electron
half-filled ring, but the transition takes place at $\phi=0$.
All the other features, viz, the continuous like variation of current with
$\phi$ and its reduction as a result of electronic correlation remain almost 
similar as presented in Fig.~\ref{f6}. But there is a crucial difference
between these two sets of electronic configurations. Comparing the 
current spectra given in Figs.~\ref{f6}(b) and \ref{f8}(b) it is seen that
the suppression of current due to e-e interaction becomes more significant
for the ring with $N_{up}=3$ and $N_{dn}=1$ compared to the other case. 
The reason is that in the first case where $N_{up} > N_{dn}$, the hopping 
probability of up spin electrons is less compared to the ring with 
$N_{up}=N_{dn}$ since in the later case pairing of up and down spin 
electrons in a single site is possible as a result of hopping term for 
\begin{figure}[ht]
{\centering \resizebox*{8cm}{8cm}{\includegraphics{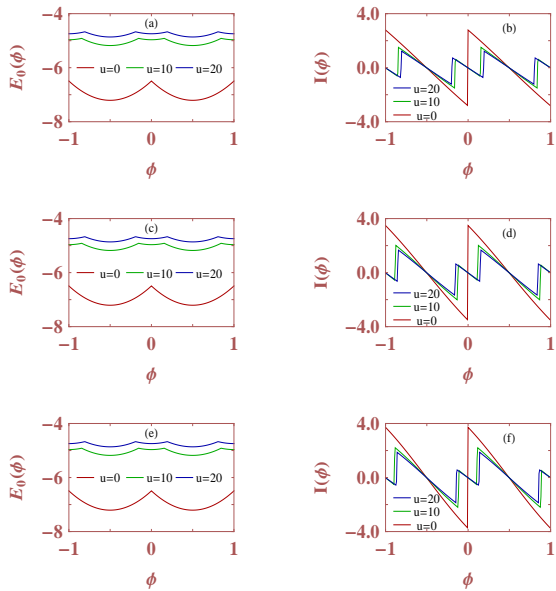}}\par}
\caption{(Color online). $E_0$-$\phi$ and $I$-$\phi$ curves for a $7$-site
ring with $N_{up}=2$ and $N_{dn}=2$ for some typical values of $U$ considering
$t_1=-0.1$ and $t_2=-0.05$, where three different rows correspond to the
identical meaning as given in Fig.~\ref{f7}.}
\label{f9}
\end{figure}
low $U$, yielding a contribution to the current. For large $U$ in both 
these two cases current gets reduced.

Focusing on the spectra given in Fig.~\ref{f9} apparently it seems that
the features are almost identical with the results shown in Fig.~\ref{f7},
but there is a striking difference between the two. For this configuration
($N_{up}=N_{dn}=2$) kink, associated with $U$-dependent states, appears 
for any non-zero correlation strength $U$, unlike the previous case
($N_{up}=3$ and $N_{dn}=1$) where a critical $U$ (say $U_c$) is required.
Apart from this, other qualitative features are almost identical. For
all these four-electron ring systems we get only $\phi_0$ periodic 
persistent current.

\vskip 0.25 cm
\noindent
\textbf{Case 4 - Ring with five electrons:} Now we move to the five-electron
ring system and here we also consider two distinct configurations depending 
on $N_{up}$ and $N_{dn}$. In one configuration we set $N_{up}=3$ and 
$N_{dn}=2$, while for the other we fix $N_{up}=4$ and $N_{dn}=1$.

\vskip 0.25 cm
\noindent
$\blacksquare$ Configuration I -- Ring with $N_{up}=3$ and $N_{dn}=2$: \\
For this configuration the many-body basis states get the forms:
$|\psi_{1}\rangle=c_{p\uparrow}^\dagger c_{q\uparrow}^\dagger c_{r\uparrow}^
\dagger c_{s\downarrow}^\dagger c_{t\downarrow}^\dagger|0\rangle$ and
$|\psi_{2}\rangle=c_{k\uparrow}^\dagger c_{l\uparrow}^\dagger c_{m\uparrow}^
\dagger c_{n\downarrow}^\dagger c_{o\downarrow}^\dagger |0\rangle$, and
the dimension of the Hamiltonian for a $N$-site ring becomes 
$(N^2/12)(N-1)^2 (N-2) \times (N^2/12)(N-1)^2 (N-2)$.

The characteristic features of ground state energy along with persistent
current for a half-filled five-electron ($N_{up}=3$ and $N_{dn}=2$) ring
\begin{figure}[ht]
{\centering \resizebox*{7cm}{5cm}{\includegraphics{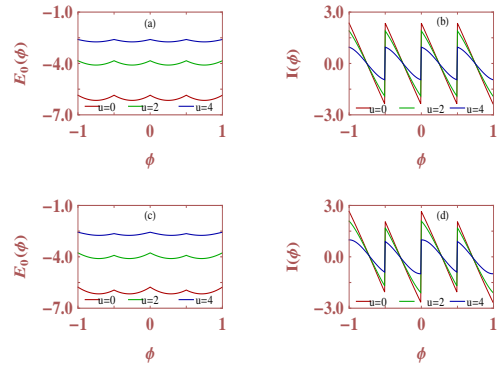}}\par}
\caption{(Color online). $E_0$-$\phi$ and $I$-$\phi$ characteristics for 
a five-electron ($N_{up}=3$ and $N_{dn}=2$) half-filled ring, where the
first row corresponds to the ring with only NNH integral and the second 
row describes the ring with NNH and SNH ($t_1=-0.1$) integrals. Here 
TNH does not make any sense.}
\label{f10}
\end{figure}
system are shown in Fig.~\ref{f10}. In the first row, results are given
when the ring is described with only NNH, while in the other row we 
consider the effect of SNH integral too. For this five-site ring inclusion
of TNH does not make any sense, since long-range hopping in shortest path
is only the physically accessible hopping parameter. Several notable 
features are obtained. (i) For the ring described with only NNH, current 
exhibits half-flux-quantum ($\phi_0/2$) periodicity and it is not at all
affected by the electronic correlation. This is exactly the same 
what we get in a three-electron half-filled ring (see Fig.~\ref{f4}).
(ii) But surprisingly we see that the $\phi_0/2$ periodicity disappears 
when the effect of higher order hopping is included (Fig.~\ref{f10}(d)).
From these results we can generalize that $\phi_0/2$ periodicity is 
specific to the odd-half-filled rings described with only NNH integral.
Other properties like the reduction of current with $U$ and the role
of higher order hopping on current enhancement remain same as discussed
for other cases.

\begin{figure}[ht]
{\centering \resizebox*{8cm}{8cm}{\includegraphics{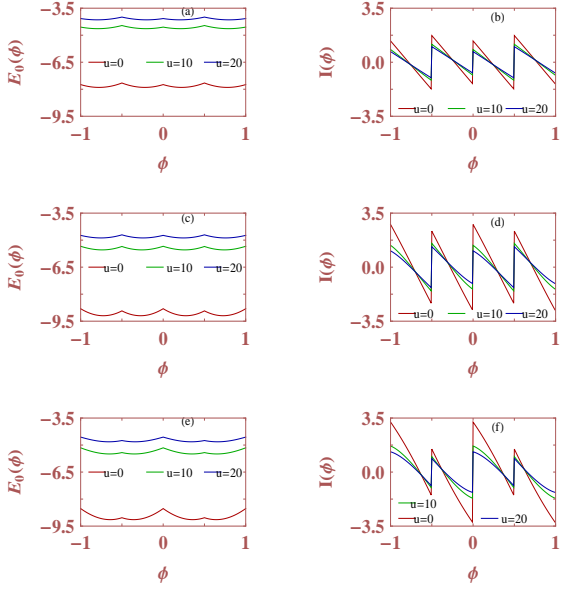}}\par}
\caption{(Color online). Variation of ground state energy and corresponding
current for a five-electron ($N_{up}=3$ and $N_{dn}=2$) non-half-filled
ring ($N=7$) for some typical values of $U$, where the $1$st, $2$nd and
$3$rd rows represent the ring with only NNH; NNH and SNH ($t_1=-0.4$);
NNH, SNH and TNH ($t_2=-0.2$) integrals, respectively.}
\label{f11}
\end{figure}

The results for a five-electron ($N_{up}=3$ and $N_{dn}=2$) non-half-filled
ring are placed in Fig.~\ref{f11} considering some typical values of $U$.
All the basic features of ground state energy and corresponding current 
shown in this spectra (Fig.~\ref{f11}) are almost similar to those given
in Fig.~\ref{f5} for a three-electron non-half-filled ring. For this 
five-electron system we can also observe kink-like structure in persistent 
current above a critical value of $U$, like the three-electron interacting 
ring. The critical value $U_c$ is very large for this $7$-site ring
(which is not shown here in the figure) and it ($U_c$) becomes smaller
for larger ring which we confirm through our detailed numerical 
calculations. The role played by higher order hopping integrals remains 
same as discussed in other non-half-filled rings and in all these cases 
current gives $\phi_0$ periodicity.

\vskip 0.25 cm
\noindent
$\blacksquare$ Configuration II -- Ring with $N_{up}=4$ and $N_{dn}=1$: \\
For this set-up the basis states read as
$|\psi_{1}\rangle=c_{p\uparrow}^\dagger c_{q\uparrow}^\dagger c_{r\uparrow}^
\dagger c_{s\uparrow}^\dagger c_{t\downarrow}^\dagger|0\rangle$ and
$|\psi_{2}\rangle=c_{k\uparrow}^\dagger c_{l\uparrow}^\dagger c_{m\uparrow}^
\dagger c_{n\uparrow}^\dagger c_{o\downarrow}^\dagger |0\rangle$. Here
the dimension of the Hamiltonian matrix is: 
$(N^2/24) (N-1) (N-2) (N-3) \times (N^2/24) (N-1) (N-2) (N-3)$ and it is
less compared to the dimension of a matrix for a $N$-site ring having 
three up and two down spin electrons.

The half-flux-quantum periodicity is still preserved for a half-filled 
\begin{figure}[ht]
{\centering \resizebox*{7cm}{5cm}{\includegraphics{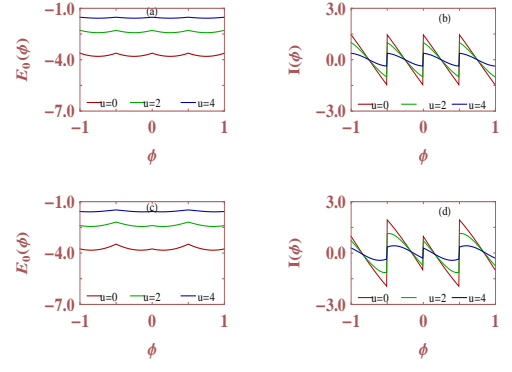}}\par}
\caption{(Color online). Ground state energy and corresponding current as
a function of $\phi$ for a half-filled five-electron ring with $N_{up}=4$
and $N_{dn}=1$, where the top row represents the ring with NNH only, and,
the bottom row describes the ring with NNH and SNH ($t_1=-0.1$) integrals.
TNH integral is redundant here.}
\label{f12}
\end{figure}
\begin{figure}[ht]
{\centering \resizebox*{8cm}{8cm}{\includegraphics{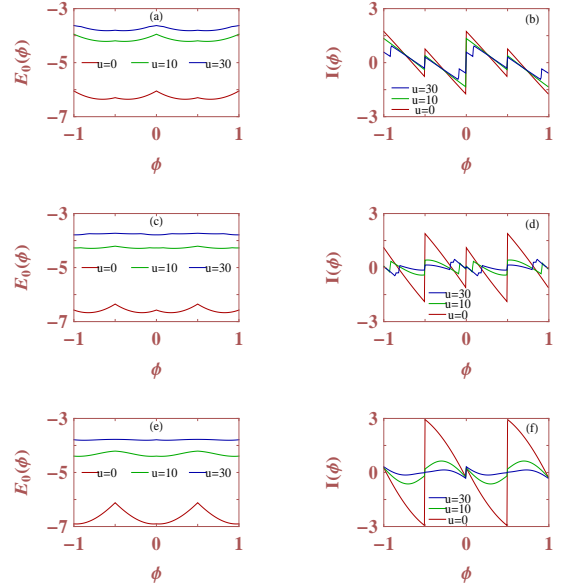}}\par}
\caption{(Color online). $E_0$-$\phi$ and $I$-$\phi$ curves for a 
five-electron ($N_{up}=4$ and $N_{dn}=1$) non-half-filled ring ($N=7$) for
some particular values of $U$, where three different rows correspond to the
identical meaning as given in Fig.~\ref{f11}.}
\label{f13}
\end{figure}
five-electron ring even when we set $N_{up}=4$ and $N_{dn}=1$. But the 
requirement is that the ring should be described with only NNH hopping, 
and the electronic correlation does not have any explicit dependence
on it. But as long as the effect of higher order hopping is incorporated, 
$\phi_0/2$ periodicity gets lost and the current regains its conventional 
$\phi_0$ periodicity. These features are clearly presented in Fig.~\ref{f12}
and consistent with the results shown in the spectra Fig.~\ref{f10}.

In the non-half-filled case some anomalous features appear in persistent
current associated with ground state energy in presence of e-e interaction.
The results for a seven-site ring are displayed in Fig.~\ref{f13} 
considering $N_{up}=4$ and $N_{dn}=1$. For $U=0$, sharp transitions are
available, while sudden phase change in current takes place at some
typical fluxes in presence of $U$ due to $U$-dependent states as seen 
from the $I$-$\phi$ spectra. Here only $\phi_0$ periodic current is 
obtained.

\vskip 0.25 cm
\noindent
\textbf{Case 5 - Ring with six electrons:} Finally, we consider six-electron 
system to study the characteristic properties of persistent current. 
Depending on up and down spin electrons we focus on two distinct electronic
configurations. For one configuration we set $N_{up}=N_{dn}=3$ and for
the other configuration we fix $N_{up}=4$ and $N_{dn}=2$.

\vskip 0.25 cm
\noindent
$\blacksquare$ Configuration I -- Ring with $N_{up}=3$ and $N_{dn}=3$: \\
The many-body basis states for this configuration look like
$|\psi_{1}\rangle=c_{p\uparrow}^\dagger c_{q\uparrow}^\dagger 
c_{r\uparrow}^\dagger c_{s\downarrow}^\dagger c_{t\downarrow}^\dagger 
c_{u\downarrow}^\dagger |0\rangle$ and
$|\psi_{2}\rangle=c_{j\uparrow}^\dagger c_{k\uparrow}^\dagger 
c_{l\uparrow}^\dagger c_{m\downarrow}^\dagger c_{n\downarrow}^\dagger 
c_{o\downarrow}^\dagger |0\rangle$, and the dimension of the Hamiltonian
matrix for a $N$-site ring becomes $(N^2/36) (N-1)^2 (N-2)^2 
\times (N^2/36) (N-1)^2 (N-2)^2$. 

\begin{figure}[ht]
{\centering \resizebox*{7cm}{5cm}{\includegraphics{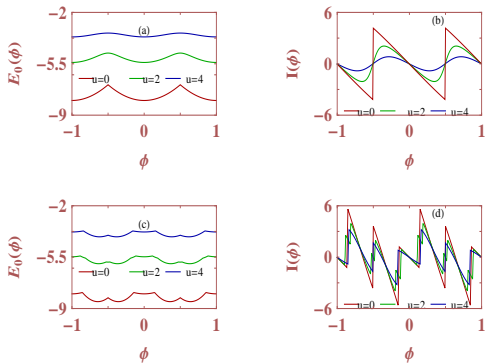}}\par}
\caption{(Color online). Ground state energy and current spectra for a
six-site ring with three up and three down spin electrons for three distinct
values of $U$, where the first row corresponds to the ring with only NNH
and the second row represents the ring with NNH and SNH integrals. Here we
set $t_1=-0.8$.}
\label{f14}
\end{figure}

In Fig.~\ref{f14} we present the dependence of ground state energy
along with persistent current for a six-site ring considering three
up and three down spin electrons for some typical values of $U$. At $U=0$ 
the current provides a sharp jump across $\phi=\pm \phi_0/2$, while it 
varies continuously with reduced amplitude in presence of finite e-e 
interaction. Exactly identical behavior is obtained for the four-electron 
half-filled ring. Thus it can be emphasized that there exists a striking 
similarity between all half-filled rings with even number of electrons 
under NNH approximation. With the addition of higher order hopping (here 
it is SNH), some anomalous oscillations appear as a result of phase 
reversal of ground state energy.

Away from half-filling kink-like structure appears following the 
variation of ground stage energy and the width of this kink decreases
\begin{figure}[ht]
{\centering \resizebox*{8cm}{8cm}{\includegraphics{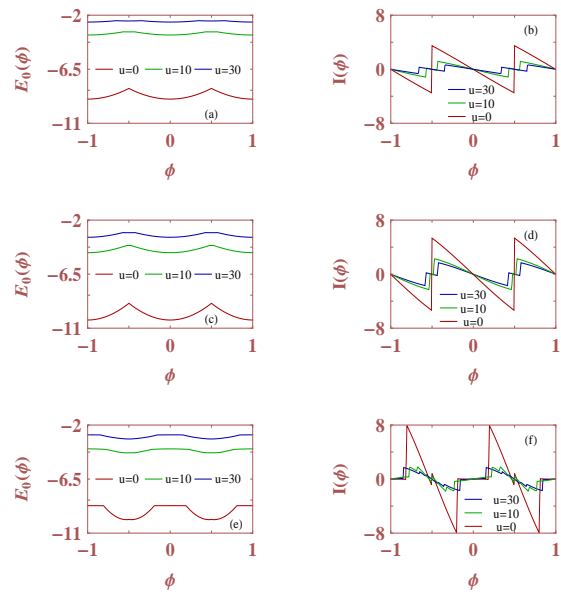}}\par}
\caption{(Color online). $E_0$-$\phi$ and $I$-$\phi$ curves for a seven-site
ring with three up and three down spin electrons, where the $1$st, $2$nd
and $3$rd rows correspond to the ring with only NNH; NNH and SNH 
($t_1=-0.6$); NNH, SNH and TNH ($t_2=-0.5$) integrals, respectively.}
\label{f15}
\end{figure}
with the incorporation of additional hopping integrals (see Fig.~\ref{f15}), 
similar to the four-electron ring system. In addition, current exhibits 
more complex structure in presence of TNH for large $U$.

\vskip 0.25 cm
\noindent
$\blacksquare$ Configuration II -- Ring with $N_{up}=4$ and $N_{dn}=2$: \\
The basis states for this choice of up and down spin electrons are:
$|\psi_{1}\rangle=c_{p\uparrow}^\dagger c_{q\uparrow}^\dagger 
c_{r\uparrow}^\dagger c_{s\uparrow}^\dagger c_{t\downarrow}^\dagger 
c_{u\downarrow}^\dagger |0\rangle$ and
$|\psi_{2}\rangle=c_{j\uparrow}^\dagger c_{k\uparrow}^\dagger 
c_{l\uparrow}^\dagger c_{m\uparrow}^\dagger c_{n\downarrow}^\dagger 
c_{o\downarrow}^\dagger |0\rangle$. This configuration sets the dimension 
of the matrix for a $N$-site ring as $(N^2/48) (N-1)^2 (N-2) (N-3) 
\times (N^2/48) (N-1)^2 (N-2) (N-3)$ and it is lower than the dimension 
of a identical size ring containing three up and three down spin electrons.

\begin{figure}[ht]
{\centering \resizebox*{7cm}{5cm}{\includegraphics{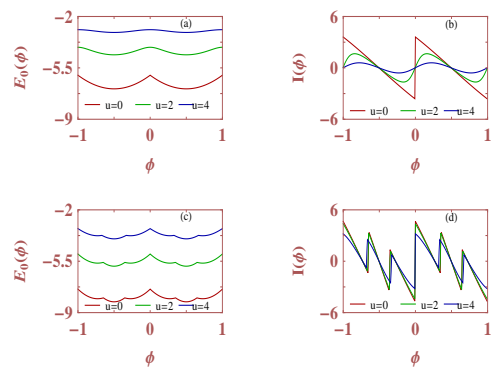}}\par}
\caption{(Color online). $E_0$-$\phi$ and $I$-$\phi$ spectra for a six-site
ring with $N_{up}=4$ and $N_{dn}=2$, where the upper and lower rows 
represent the similar meaning as described in Fig.~\ref{f14} with identical
parameter values.}
\label{f16}
\end{figure}

The characteristics features of ground state energy together with persistent
current for a half-filled six-electron ring with $N_{up}=4$ and $N_{dn}=2$
are shown in Fig.~\ref{f16} for some typical values of $U$. Comparing the
spectra given in Figs.~\ref{f14} and \ref{f16} it is noticed that almost
identical variation of current is obtained apart from a shifting along 
the flux line. In one case the sharp transition for $U=0$ takes place 
at $\phi=\pm \phi_0/2$ (Fig.~\ref{f14}), while in the other case it occurs 
at $\phi=0$ (Fig.~\ref{f16}) for the identical strength of $U$. These 
features, in fact, are quite similar to the results discussed for two 
different configurations in the half-filled four-electron ring system. 
Certainly, these observations demand that all half-filled rings with even 
number of electrons exhibit qualitatively identical patterns.

Quite similar argument can also be applicable for the non-half-filled case.
\begin{figure}[ht]
{\centering \resizebox*{8cm}{8cm}{\includegraphics{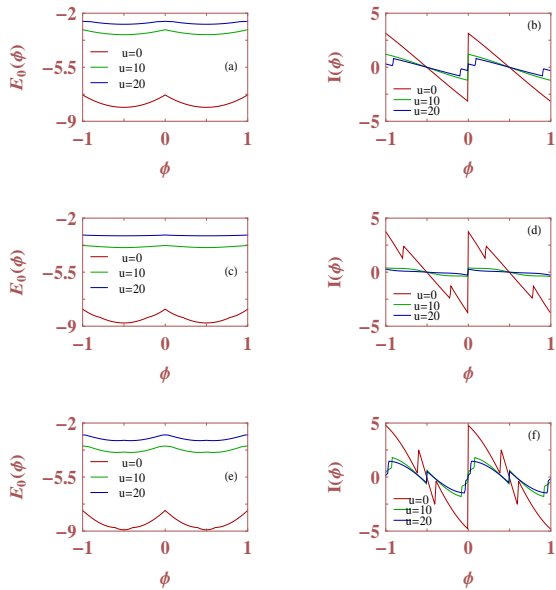}}\par}
\caption{(Color online). Dependence of ground state energy and associated
current with flux $\phi$ for a seven-site ring considering $N_{up}=4$ and 
$N_{dn}=2$, where different rows correspond to the identical meaning as 
stated in Fig.~\ref{f15}. Here we set $t_1=-0.5$ and $t_2=-0.4$.}
\label{f17}
\end{figure}
Apart from certain kinks at different fluxes, current exhibit more
complex structures in presence of higher order hopping as clearly seen
from the spectra given in Fig.~\ref{f17}, like the non-half-filled ring 
with three up and three down spin electrons (see Fig.~\ref{f15}). Due to
smaller size of the ring (where $N_e$ is very close to $N$) current gets
suppressed significantly with increasing the correlation strength and
this effect dominates for the maximum unequal distribution among up and 
down spin electrons keeping the total number of electrons constant.
For all these six-electron ring systems we get only $\phi_0$ periodic 
persistent current, like two-electron and four-electron cases.

\subsubsection{Effect of impurity} 

Now we investigate the role of impurities on persistent current and the
interplay between Hubbard interaction and long-range hopping on it. To 
introduce impurities in a ring we choose on-site potentials ($\epsilon_j$)
randomly from a `Box' distribution function of width $W$ setting the 
window region $-W/2$ to $W/2$. Since the response strongly depends on
disordered configurations, we compute all the results taking average 
over a large number of distinct such configurations and below we present
some of these essential results.

\begin{figure}[ht]
{\centering \resizebox*{6cm}{5.5cm}{\includegraphics{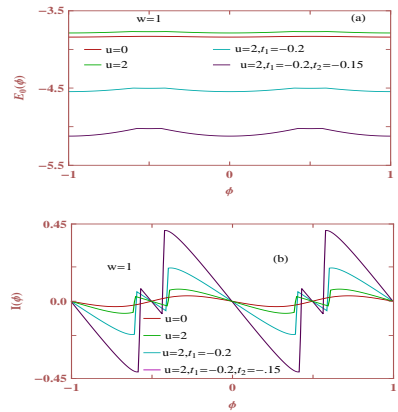}}\par}
\caption{(Color online). $E_0$-$\phi$ and $I$-$\phi$ characteristics for 
a $20$-site ring ($N=20$) with two-opposite spin electrons in presence of 
disorder ($W=1$) for some typical values $U$ considering the effect of 
higher order hopping integrals.}
\label{f18}
\end{figure}

\vskip 0.25 cm
\noindent
\textbf{Case I - Ring with two electrons:} To investigate the critical role
played by disorder in Fig.~\ref{f18} we display the dependence of ground 
state energy together with persistent current as a function of $\phi$ for 
a twenty-site
ring considering two opposite spin electrons setting the disorder strength
$W=1$. In presence of disorder current becomes vanishingly small (red curve
of Fig.~\ref{f18}(b)) for the non-interacting ring ($U=0$) described with
NNH integral. This is solely due to electronic localization as electrons get
pinned at different sites associated with the on-site potentials, and, it
becomes even smaller for higher $W$ (not shown here in the figure to save
space). Now under this situation current gets enhanced if we include the 
effect of electronic correlation(green line of Fig.~\ref{f18}(b)). This 
correlation tries to homogenize the system providing enhanced current.
But the enhancement becomes more significant when the contributions from
higher order hopping integrals are taken into account. This is clearly 
seen from the cyan and purple curves of Fig.~\ref{f18}(b). Thus it can be 
emphasized that the effect of localization as a result of disorder in NNH
model can be suppressed significantly by introducing higher order hopping
integrals in presence of e-e interaction. The available current for the
disordered ring in presence of higher order hopping integrals is quite 
comparable to the current obtained for a {\em perfect} non-interacting 
ring described with only NNH integral which, on the other hand, is very 
close to the experimental observations. Therefore, only NNH model is 
{\em not sufficient} to verify experimental results where the rings 
invariably contain disorder and certainly we have to go beyond NNH 
model.

Apart from this we also find that sudden phase reversal of current 
takes place around $\phi=\pm \phi_0/2$ due to $U$-independent states,
like the two-electron ordered ring, but for this disordered case the kink
appears beyond a critical value of $U$ ($=U_c$, say) and it depends on
ring size as well as disorder strength. Whereas, any non-zero value of 
$U$ yields kink-like structure in the ordered ring. For all these cases
current exhibits $\phi_0$ flux-quantum periodicity.

\vskip 0.25 cm
\noindent
\textbf{Case II - Ring with three electrons:} Finally, let us focus
on the response of a three-electron disordered ring. 

\begin{figure}[ht]
{\centering \resizebox*{6cm}{6cm}{\includegraphics{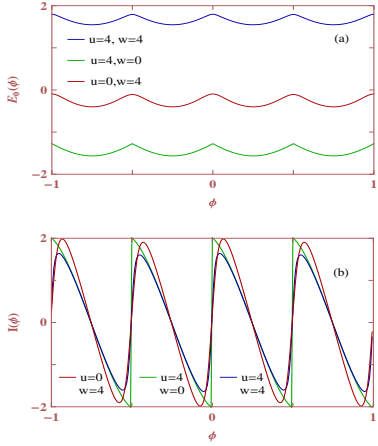}}\par}
\caption{(Color online). $E_0$-$\phi$ and $I$-$\phi$ characteristics 
for a three-electron ($N_{up}=2$ and $N_{dn}=1$) ring in the half-filled 
case in presence of disorder. For the sake of comparison, the result of
an ordered ring (green line) is also given.}
\label{f19}
\end{figure}

In Fig.~\ref{f19} we present the variation of ground state energy together
with persistent current for a half-filled ($N=3$) disordered ring,
described with NNH integral, considering $N_{up}=2$ and $N_{dn}=1$. 
To reveal the precise role of disorder on periodicity of ground state
energy as well as current with $\phi$, here we superimpose the result of
an ordered interacting ring (green curve). From the spectra it is seen 
that $\phi_0/2$ periodicity no longer exists in presence of disorder,
and both the energy and current get usual $\phi_0$ periodicity.

For this three-site ring it is quite hard to view the competition between
\begin{figure}[ht]
{\centering \resizebox*{6cm}{6cm}{\includegraphics{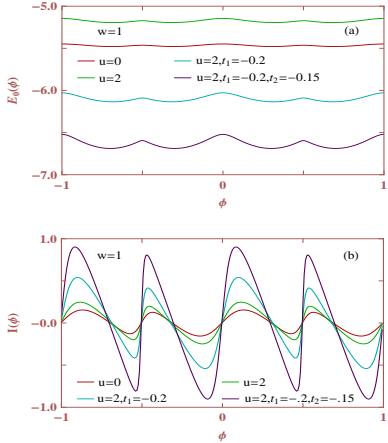}}\par}
\caption{(Color online). Ground state energy and corresponding current for 
a three-electron ($N_{up}=2$ and $N_{dn}=1$) ring in the non-half-filled
case ($N=10$) in presence of disorder.}
\label{f20}
\end{figure}
$U$ and $W$, and thus, we focus on the spectra given in Fig.~\ref{f20} 
where the results are shown for a ten-site ring. A significant enhancement
of current takes place in presence of electronic correlation and higher 
order hopping integrals, and, sometimes this enhancement becomes two
orders of magnitude larger compared to the NNH model.

Before we finish this sub-section we would like to point out that the
characteristic features of ground state energies and corresponding 
persistent currents for other disordered rings with four, five and even
six electrons are reviewed thoroughly and from our analysis we find exactly
similar kind of enhancement and identical periodic nature of current in 
the presence of e-e interaction and long-range hopping. 

\vskip 0.5cm
\begin{center}
{\bf Scaling behavior}
\end{center}
\vskip 0.3cm
The results analyzed so far are worked out for some typical interacting
rings taking different combinations of up and down spin electrons, and,
these results exhibit certain similarities among the rings with even
number (two, four and six) of electrons as well as rings with odd number
(three and five) of electrons. To make the analysis a self-contained one,
now we discuss the dependence of current with ring size $N$ and from it
we establish its scaling behavior. 

\begin{figure}[ht]
{\centering \resizebox*{6.5cm}{6.25cm}{\includegraphics{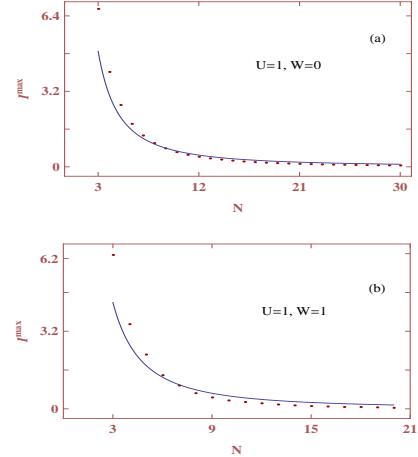}}\par}
\caption{(Color online). Dependence of persistent current with ring size
$N$ considering $U=1$ and $N_e=2$ ($N_{up}=N_{dn}=1$). The dotted points, 
corresponding to the currents, are evaluated following the exact 
diagonalization method. Using these dots a scaling relation between the 
current and ring size is established which generates the continuous curve.
Two cases are presented depending on $W$ where (a) and (b) correspond to
$W=0$ and $1$, respectively.}
\label{scl1}
\end{figure}

Figure~\ref{scl1} displays the dependence of current ($I^{\mbox{\tiny max}}$)
as a function of $N$ considering $U=1$ and $N_e=2$ ($N_{up}=N_{dn}=1$), where 
$I^{\mbox{\tiny max}}$ is obtained by taking the maximum absolute value of 
current computed within the flux range $0$ to $\phi_0$. Two different cases, 
one is for $W=0$ and other is for $W=1$, are described those are given in 
(a) and (b), respectively. The dotted points, corresponding to current 
for different 
$N$, are computed following the exact diagonalization method, and, using 
these dots we construct a scaling behavior of current which generates the 
continuous line. The scaling relation gets the form: 
$I^{\mbox{\tiny max}}=CN^{-\xi}$, where the constant $C$ becomes $30$ for 
both ordered and disordered cases, while the exponent $\xi$ becomes different 
in these two cases. It is $1.65$ for $W=0$ and $1.75$ for $W=1$. Usually, 
for a particular filling this constant ($C$) depends on the disorder strength, 
but for this typical $N_e$ we find its
\begin{figure}[ht]
{\centering \resizebox*{6.5cm}{6.25cm}{\includegraphics{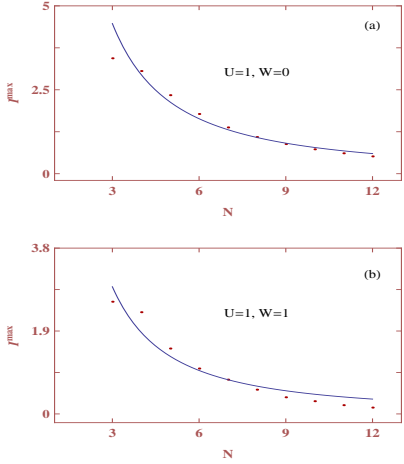}}\par}
\caption{(Color online). Same as Fig.~\ref{scl1} with $N_e=3$, where 
$N_{up}=2$ and $N_{dn}=1$.}
\label{scl2}
\end{figure}
identical value both for the ordered and disordered rings which we confirm 
through elaborate numerical calculations. 

In Fig.~\ref{scl2} we present the variation of current with ring size $N$
for $N_e=3$ (considering $N_{up}=2$ and $N_{dn}=1$), keeping all other 
physical parameters the same as taken in Fig.~\ref{scl2}. Like two-electron
system, here we also get similar kind of scaling relation, viz, 
$I^{\mbox{\tiny max}}=CN^{-\xi}$, where the constant $C$ becomes $22$ and
$16$, and, the exponent $\xi$ reaches to $1.45$ and $1.55$ for the ordered
and disordered cases, respectively.

In the same way we can also check scaling behavior of persistent current
in higher-electron systems (i.e., rings with four, five, six and more
electrons), and from our analysis it can be emphasized that in each case
current obeys the identical scaling relation, viz, $I \propto N^{-\xi}$.
This exponent $\xi$ gradually decreases, both for ordered and disordered 
cases, with $N_e$ and for $N_e \geq 5$ it almost reaches to the limiting 
value $1.33$ at $U=1$, which is consistent with the previous analysis
done by Gendiar {\em et al}~\cite{gen} using bosonization techniques.
Our analysis clearly suggests that the current decreases algebraically
with ring size $N$.

\subsection{Hartree-Fock mean field analysis}

In this sub-section we present numerical results computed from HF mean-field 
approach and compare these results with those obtained from the other method 
i.e., exact numerical diagonalization.

\begin{figure}[ht]
{\centering \resizebox*{7cm}{5.5cm}{\includegraphics{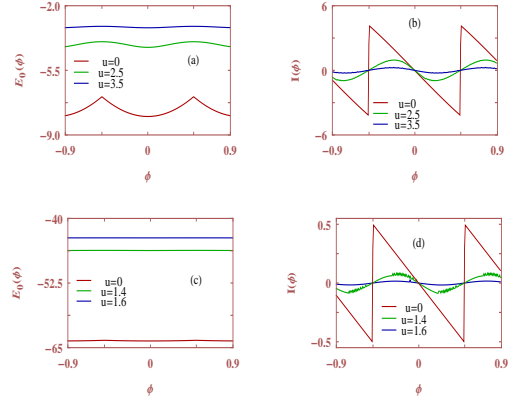}}\par}
\caption{(Color online). Dependence of ground state energy and corresponding
persistent current as a function of $\phi$ for some typical ordered rings,
described with only NNH integral, in the half-filled band case considering 
different values of $U$. In the first row we choose $N=6$ and
$N_{up}=N_{dn}=3$, while in the second row these parameters are: $N=50$
and $N_{up}=N_{dn}=25$.}
\label{mf1}
\end{figure}

Figure~\ref{mf1} displays the dependence of ground state energy together
with persistent current as a function of flux $\phi$ for some typical 
ordered rings (even $N$) described with only NNH integral in the limit 
of half-filling
considering different values of electronic correlation strength $U$. Two
cases are analyzed depending on the ring size $N$. In the first row we 
present the results for $N=6$ and $N_{up}=N_{dn}=3$, to compare these
spectra with our previous data (top row of Fig.~\ref{f14}) obtained from 
\begin{figure}[ht]
{\centering \resizebox*{7cm}{5.5cm}{\includegraphics{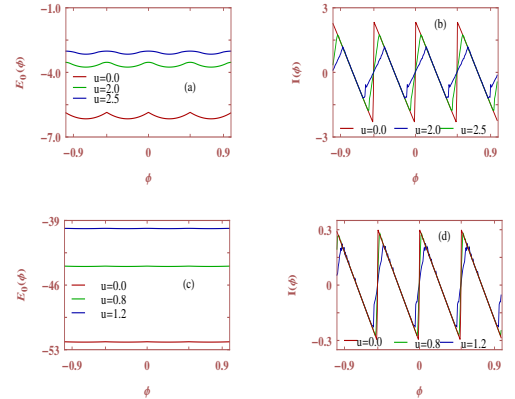}}\par}
\caption{(Color online). $E_0$-$\phi$ and $I$-$\phi$ spectra for some typical 
ordered rings, described with only NNH integral, in the half-filled band 
case considering different values of $U$. In the first row we choose $N=5$, 
$N_{up}=3$ and $N_{dn}=2$, while in the second row these parameters are: 
$N=41$, $N_{up}=21$ and $N_{dn}=20$.}
\label{mf2}
\end{figure}
exact numerical diagonalization method. Comparing the spectra 
Figs.~\ref{mf1}(b) and \ref{f14}(b) we see that MF results are highly 
consistent with our data obtained from exact numerical diagonalization 
of the many-body Hamiltonian. Therefore, we can certainly rely on MF 
solutions and extend our analysis to larger lings. As illustrative example,
in the second row of Fig.~\ref{mf1} we present the MF results for 
a $50$-site ring, and they exhibit almost identical features, like a
$6$-site ring, which yield the invariant nature of energy- and current-flux 
characteristics for different ring sizes when the filling factor remains 
constant.

In a similar way we now focus our attention on the rings with odd $N$.
The results are shown in Fig.~\ref{mf2} where we present the behavior of 
ground state energy and corresponding persistent current, obtained from MF 
technique, for two different ring sizes. The spectra for $N=5$ (first row
of Fig.~\ref{mf2}) can be directly compared to the results given in the
first row of Fig.~\ref{f10} since all the physical parameters are same 
for these two cases. Quite interestingly we see that almost all the 
features, viz, suppression of current with $U$ and half flux-quantum 
periodicity as a function of $\phi$ remain unchanged. But, for narrow
regions of $\phi$ the slope of the currents for non-zero $U$ shows opposite
signature in these two different numerical methods, while it (slope) remains
\begin{figure}[ht]
{\centering \resizebox*{6.6cm}{6cm}{\includegraphics{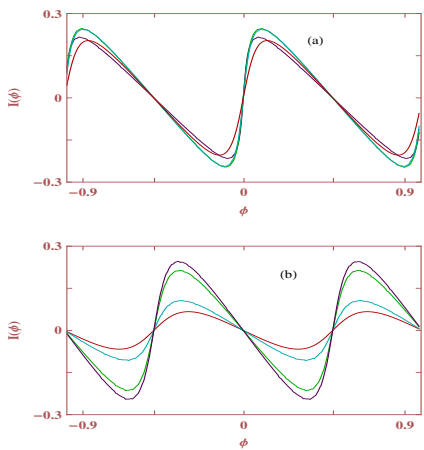}}\par}
\caption{(Color online). Variation of persistent current for a $60$-site 
disordered ($W=1$) ring, where (a) $N_{up}=N_{dn}=30$ 
and (b) $N_{up}=N_{dn}=15$. The red line corresponds to the disordered 
non-interacting ($U=0$) ring with only NNH integral, while introducing the
effect of e-e interaction ($U=0.5$) in this ring we get cyan curve. The 
green curve represents the ring with disorder, e-e interaction ($U=0.5$)
and SNH integral ($t_1=-0.2$), whereas the purple line corresponds to the 
ring with disorder, e-e interaction ($U=0.5$), SNH ($t_1=-0.2$) and TNH
($t_2=-0.05$) integrals.}
\label{mf3}
\end{figure}
same for other flux regions. This is the peculiarity of MF calculations
and can be noticed for any other ordered rings with odd $N$ in the limit
of half-filling. To substantiate it in the second row of Fig.~\ref{mf2}
we present the results for a $41$-site ring considering $N_{up}=21$
and $N_{dn}=20$, and see that all the properties exhibited by a $5$-site
ring are nicely followed in this bigger ring.

Finally, let us focus on the MF results given in Fig.~\ref{mf3} where we 
present the variation of persistent current for a $60$-site ring as a 
function of $\phi$ considering two different band fillings those are placed
in (a) and (b), respectively. Four distinct cases are analyzed depending on 
the physical parameters of the system. The red curve corresponds to the
current for the non-interacting ($U=0$) disordered ($W=1$) ring described
with only NNH integral. While introducing the effect of e-e interaction in 
this ring we get cyan curve. The green line represents the current for the 
ring with disorder, e-e interaction and SNH integral, whereas incorporating 
further the effect of TNH integral in this ring we get the current shown by
the purple curve. At half-filling all the currents are quite comparable with 
each other (Fig.~\ref{mf3}(a)) and higher order hopping integrals do not 
actually play any such significant role in enhancing the current since it 
is counterbalanced by the repulsive Coulomb interaction. On the other hand, 
at less than half-filling an enhancement of current takes place as a result 
of $U$ and it becomes more significant in presence of higher order hopping 
integrals (Fig.~\ref{mf3}(b)) as there exists empty sites where electrons 
can hop quite easily yielding more current. These mean-field results are 
fully consistent with our exact numerical diagonalization analysis (see 
Figs.~\ref{f19} and \ref{f20}).

\section{Closing Remarks}

To conclude, in this work we have investigated the behavior of ground state
energy and persistent current in interacting mesoscopic rings subjected to
AB flux $\phi$ in presence of long-range hopping and disorder. Two different
methods, exact numerical diagonalization and HF mean-field theory, have been
used to compute numerical results of the many-body systems, and, found that
the results obtained from these two methods are highly consistent with each
other. The interplay between Hubbard interaction and long-range hopping
integrals yields a significant enhancement of persistent current in 
disordered ring which in some cases becomes comparable to that of an 
ordered ring and reaches very close to the experimental observations. 
This phenomenon suggests that only e-e interaction is not sufficient to
explain the enhancement of persistent current in a disordered ring described 
with NNH integral. We have to incorporate the effect of higher order hopping
integrals. From our results we have shown that even for too small values 
of $t_1$ and $t_2$ compared to $t$, current gets enhanced significantly.
In addition, we have also discussed the appearance of $\phi_0/2$ periodic
current for some typical electron fillings and its sensitivity on higher
order hopping as well as randomness. Finally, we have analyzed the scaling 
behavior of current which suggests an asymptotic decay with ring size $N$.

\end{document}